\documentclass[a4paper,11pt]{article}

\usepackage{pos}
\usepackage{graphicx}
\usepackage{caption}
\usepackage{subcaption}
\usepackage[per-mode=symbol]{siunitx}
\usepackage{wrapfig}
\usepackage{cleveref}
\usepackage{xspace}

\newcommand{\refref}[1]{Ref.~\cite{#1}}

\newcommand{\reffig}[1]{Fig.~\ref{#1}}
\newcommand{\refsec}[1]{Section~\ref{#1}}
\newcommand{\refeq}[1]{Eq.~(\ref{#1})}

\newcommand{\xmax}{\ensuremath{X_{\rm max}}\xspace}
\newcommand{\chisquared}{\ensuremath{\chi^2}\xspace}

\title{Estimation of $X_\mathrm{max}$ for air showers measured at IceCube with elevated radio antennas of a prototype surface station}

 \ShortTitle{\xmax{} estimate with radio antennas of IceCube Surface Enhancement prototype station}

\author{The IceCube Collaboration \\{\normalsize \normalfont(a complete list of authors can be found at the end of the proceedings)}\\}

\emailAdd{rturcotte@icecube.wisc.edu}
\emailAdd{verpoest@udel.edu}
\emailAdd{megha.venugopal@kit.edu}

\abstract{
The IceCube Neutrino Observatory at the geographic South Pole is, with its surface and in-ice detectors, used for both neutrino and cosmic-ray physics. The surface array, named IceTop, consists of ice-Cherenkov tanks grouped in 81 pairs spanning a \SI{1}{\km\squared} area. 
An enhancement of the surface array, composed of elevated scintillation panels and radio antennas, was designed over the last years in order to increase the scientific capabilities of IceTop. The surface radio antennas, in particular, will be able to reconstruct $X_\mathrm{max}$, an observable widely used to determine the mass composition of cosmic rays. 
A complete prototype station of this enhanced array was deployed in the Austral summer of 2019/20 at the South Pole. This station comprises three antennas and eight scintillation panels, arranged in a three-arms star shape. The nominal frequency band of the radio antennas is 70 to 350\,MHz.

In this work, we use a state-of-the-art reconstruction method in which observed events are compared directly to CoREAS simulations to obtain an estimation of the air-shower variables, in particular, energy and \xmax{}. We will show the results in this unique frequency band using the three prototype antennas.

\vspace{4mm}
{\bfseries Corresponding authors:}
Roxanne Turcotte$^{1*}$, Stef Verpoest$^{2}$, Megha Venugopal$^{1}$\\
{$^{1}$ \itshape Karlsruhe Institute of Technology, Institute for Astroparticle Physics, 76021 Karlsruhe, Germany}\\
{$^{2}$ \itshape Bartol Research Institute and Dept. of Physics and Astronomy, University of Delaware, Newark, DE 19716, USA}\\[4mm]
$^*$ Presenter

\ConferenceLogo{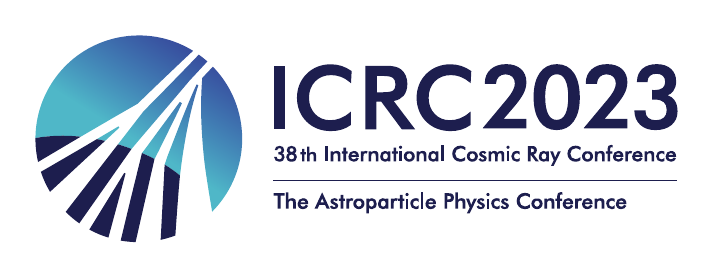}

\FullConference{The 38th International Cosmic Ray Conference (ICRC2023)\\ 26 July -- 3 August, 2023\\ Nagoya, Japan}
}

\begin{document}
\maketitle

\section{Introduction}

The IceCube Neutrino Observatory is a multi-purpose particle detector located at the geographic South Pole. Its surface component IceTop is used to study air showers from cosmic rays (CR) with energies between approx.~\SI{1}{\peta\eV} and \SI{1}{\exa\eV}~\cite{IceTopSpectrum}. An enhancement of the surface array with scintillator panels and radio antennas has been proposed to deal with the issue of snow accumulation on IceTop and to improve the identification of the properties of the primary cosmic rays~\cite{Haungs:2019ylq}.

The proposed surface array is designed to consist of 32 stations distributed across a \SI{1}{\km\squared} area, each equipped with 8 scintillation panels and three radio antennas, where the scintillation panels trigger the radio antennas. The current requirement for radio triggering is fulfilled when six panels observe signals above threshold within 1\textmu s.
In January 2020, a prototype station was deployed, measuring CR-induced air showers and enabling coincident measurements between the radio antennas and the Cherenkov tanks of IceTop. More details about the prototype station can be found in \refref{TurcotteICRC21}, while information about the observed air showers are available in Refs.~\cite{HrvojeICRC21,AbdulICRC23}. In January 2022, the Data Acquisition System (DAQ), known as TAXI 3.0~\cite{TurcotteICRC21}, was upgraded to its successor, TAXI 3.2~\cite{thesisRox}.
In this study, we employ a state-of-the-art method, first described in \refref{LOFAROriginal}, for the reconstruction of the depth of shower maximum (\xmax{}). We apply this method with different filtering schemes to analyze air showers recorded during the operation of both TAXI 3.0 and TAXI 3.2.

\section{The method}

A template-fitting method is employed in this study to compare the two-dimensional lateral distribution function (2D-LDF) of the radio signal from air-shower simulations using CORSIKA~\cite{CORSIKA} to the measured air-shower data. For each event recorded with the radio antennas, a specific set of simulations is generated to match the reconstructed trajectory and estimated energy of that particular event. 
The comparison between the measured and simulated waveforms is performed using a chi-squared (\chisquared{}) minimization method, inspired by similar techniques employed in other experiments such as \cite{LOFARTemplateLatest,Bezyazeekov:2018yjw,TemplateAERA}. In this work, the two polarization channels of the antennas are treated independently, resulting in a minimization process performed on six data points for each event. The measured signal is compared to the corresponding simulated signal using the following equation:

\begin{equation}
\label{eq:chi2}
\chi^2 = \sum^{6}_{i=1} \left( \frac{\varepsilon_i - f\cdot \varepsilon_{i,\,{\rm MC}}}{\sigma_{i}}\right)^2,
\end{equation}

where $i$ denotes the antenna channel, $\varepsilon_i$ is the maximum of the Hilbert envelope extracted from the measured waveform, $\varepsilon_{\text{MC}_i}$ is the corresponding quantity from the simulated waveform, $\sigma_i$ represents the noise contribution, and $f$ is the free parameter in the fit. The noise contribution $\sigma_i$ is determined by calculating the root mean square (RMS) in multiple time windows of 64\,ns from the measured waveform $i$, see~\refref{thesisRox}. The parameter $f$, which is the parameter to be fit, accounts for energy estimation and calibration uncertainties. 

Previous work has shown that with more than three antennas, a more detailed log-likelihood approach enhances the resolution on \xmax~\cite{TurcotteARENA}. The focus of this work is primarily on improving the filtering schemes and comparing the data acquisition systems (DAQ) used. Therefore, the simpler \chisquared{} method of \cref{eq:chi2} is used.
Furthermore, due to the limited number of antennas,  the core position reconstructed by IceTop is used. Once the array is larger, it will be possible, by using a star-pattern simulation and extrapolation method~\cite{Radcube} to add the core as a parameter to fit in \cref{eq:chi2}.

\begin{figure}
    \centering
    \includegraphics[width=0.49\linewidth]{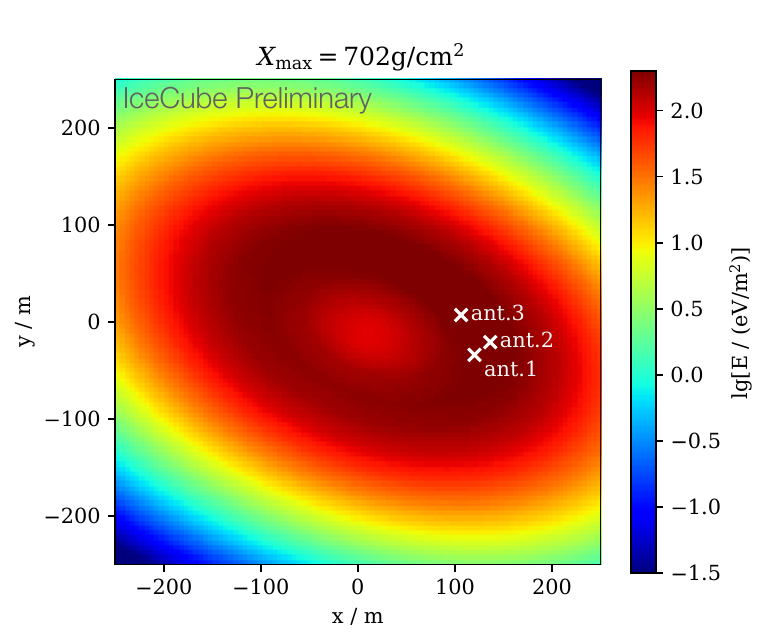}
    \includegraphics[width=0.49\linewidth]{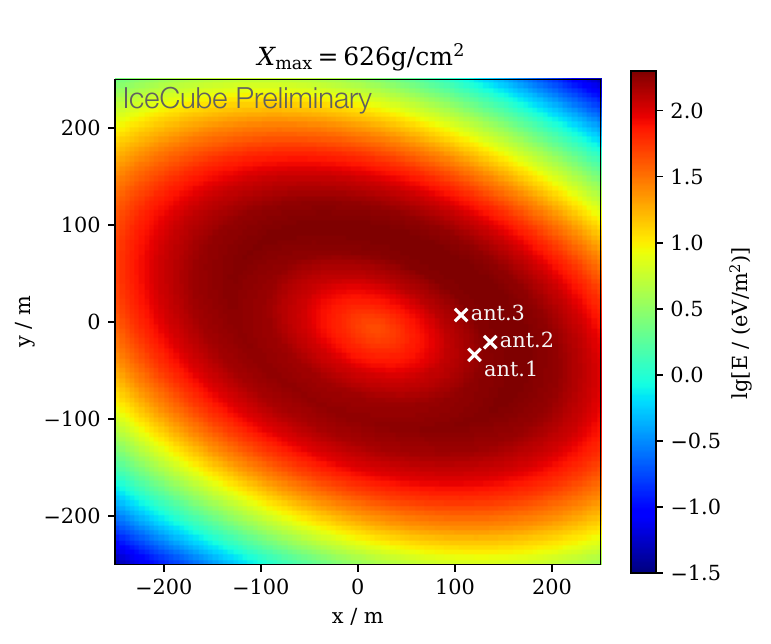}
    \caption{Radiation footprint of two simulated air showers with identical primary cosmic rays, with differing \xmax{}. The white crosses show the location of the prototype station antennas. Footprint obtained by interpolating CoREAS simulation using the method from \refref{Radcube}.}
    \label{fig:footprint}
\end{figure}

\reffig{fig:footprint} illustrates the footprint of two simulations of one air shower where the only difference is their \xmax{}. 
The figure showcases the limited prototype station area in comparison to the radiation footprint. Additionally, it demonstrates the difference in the footprint resulting from variations in the \xmax of the air-shower simulation.

\section{Dataset and processing}
\label{sec:dataset_processing}

Two different sets of recorded air showers are used in this work, which will be referred to as \textit{Set 1} and \textit{Set 2}. Both sets consist of prototype station events that have a coincident trigger in IceTop. The identification of air showers in Set 1 can be found in \refref{HrvojeICRC21}, while Set 2 is documented in \refref{AbdulICRC23}. It is important to mention that Set 1 has air showers measured with TAXI 3.0, while Set 2 only has air showers measured with TAXI 3.2. Importantly, also the method for identifying air showers differs between the two sets; the second set uses machine learning techniques to identify signals, enabling a selection of lower-energy showers. Since for both sets all of the air showers are in coincidence with an IceTop trigger, the reconstruction of the core position, direction, and shower size are derived from the IceTop detector, which is much larger and better studied compared to the currently deployed set of scintillation panels. For the initial energy estimate of the shower, a custom conversion from the reconstructed IceTop shower-size parameter, $S_{125}$~\cite{IceTop}, is implemented, as most of the showers fall outside of the commonly-used IceTop phase space.

The measured data is processed using a beamforming technique~\cite{LOPES:2021ipp}, which consists of shifting the waveforms in the different antennas in the time domain based on the arrival direction of the shower reconstructed with IceTop. The squares of the different waveforms are summed and the maximum value is found to identify the radio pulse. The maximum of the Hilbert envelope is then calculated in a 50\,ns window around the pulse location and used as the observable ($\varepsilon$) in \refeq{eq:chi2}.  

For each recorded air-shower event, $\sim$75 air showers are simulated for those in Set 1, and $\sim$45 for those in Set 2, using both protons and iron nuclei as primary cosmic rays. CORSIKA~\cite{CORSIKA} is used for simulating the air showers, while CoREAS~\cite{CoREAS} is employed to simulate the radio emission. The instrumental response is incorporated into the simulations using a dedicated framework within the IceCube software specifically designed for radio data processing~\cite{Radcube}. The simulated electric field is transformed into measurements of the two polarization channels of each antenna, and the resulting waveforms are filtered between 80-300\,MHz.

\begin{figure}
    \centering
    \includegraphics[width=0.99\textwidth]{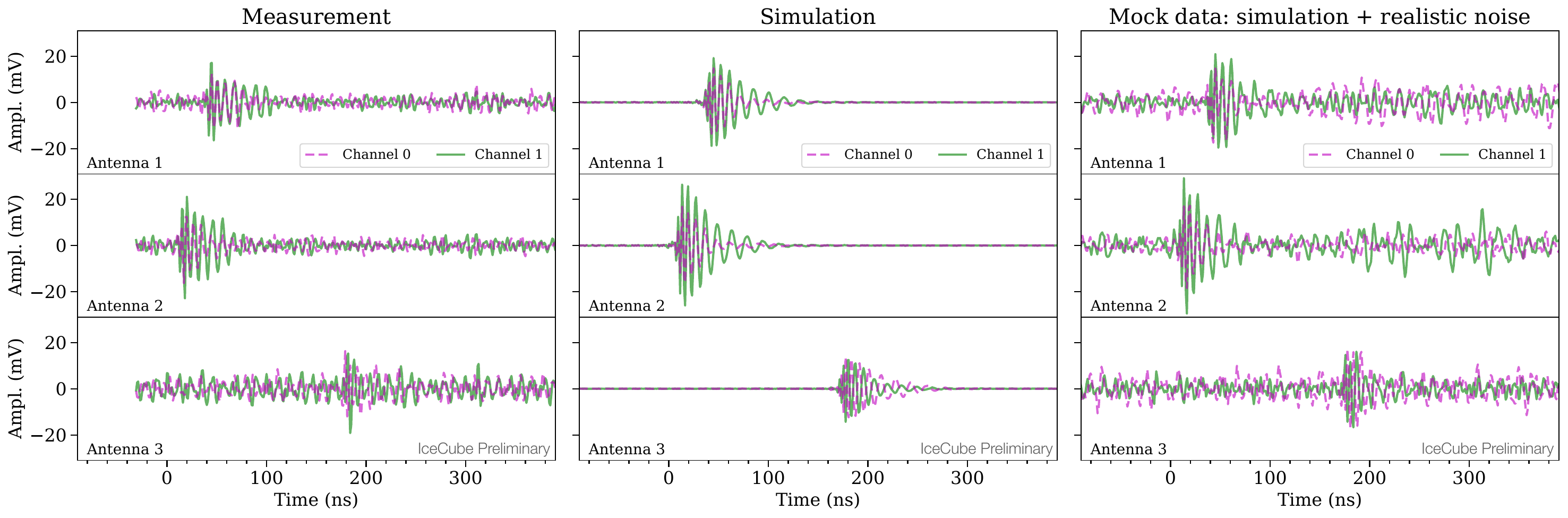}
    \caption{The waveform from the three antennas after undergoing the processing pipeline and being filtered between 80-\SI{300}{\mega\Hz}. The leftmost waveform shows measured waveforms, the middle waveform shows an example of the signal from a simulated event, and the rightmost waveform represents the same simulated waveform with measured noise injected.}
    \label{fig:waveforms}
\end{figure}

\reffig{fig:waveforms} gives an example of the waveforms for one event. The left panel displays the measured waveforms from the event. The middle panel displays the waveforms from one of the simulations performed for this specific event after the processing pipeline, without the addition of any noise. The right panel depicts the simulated waveform (from the middle panel) with real background noise injected into the traces. Here, "real noise" refers to noise obtained from background waveforms recorded on the day of the event. For the application of the technique with mock data (\refsec{sec:mock_data}) or for the analysis of real data (\refsec{sec:data}), the quantity $\varepsilon_{i}$ is sourced from the right panel (simulated waveform with injected noise) or the left panel (measured waveform) respectively, and in both case is compared to the middle panel ($\varepsilon_{\mathrm{MC}_i})$.

\section{Reconstruction estimation with mock data}
\label{sec:mock_data}

The application of the reconstruction technique to mock data serves two purposes: to validate the reconstruction method and to estimate the resolution of different methodologies. The process is straightforward: real noise is added to the simulations, turning them into mock data, which are then compared to the noiseless simulations as if they were measured data.

\begin{figure}
    \centering
    \includegraphics[width=0.85\linewidth]{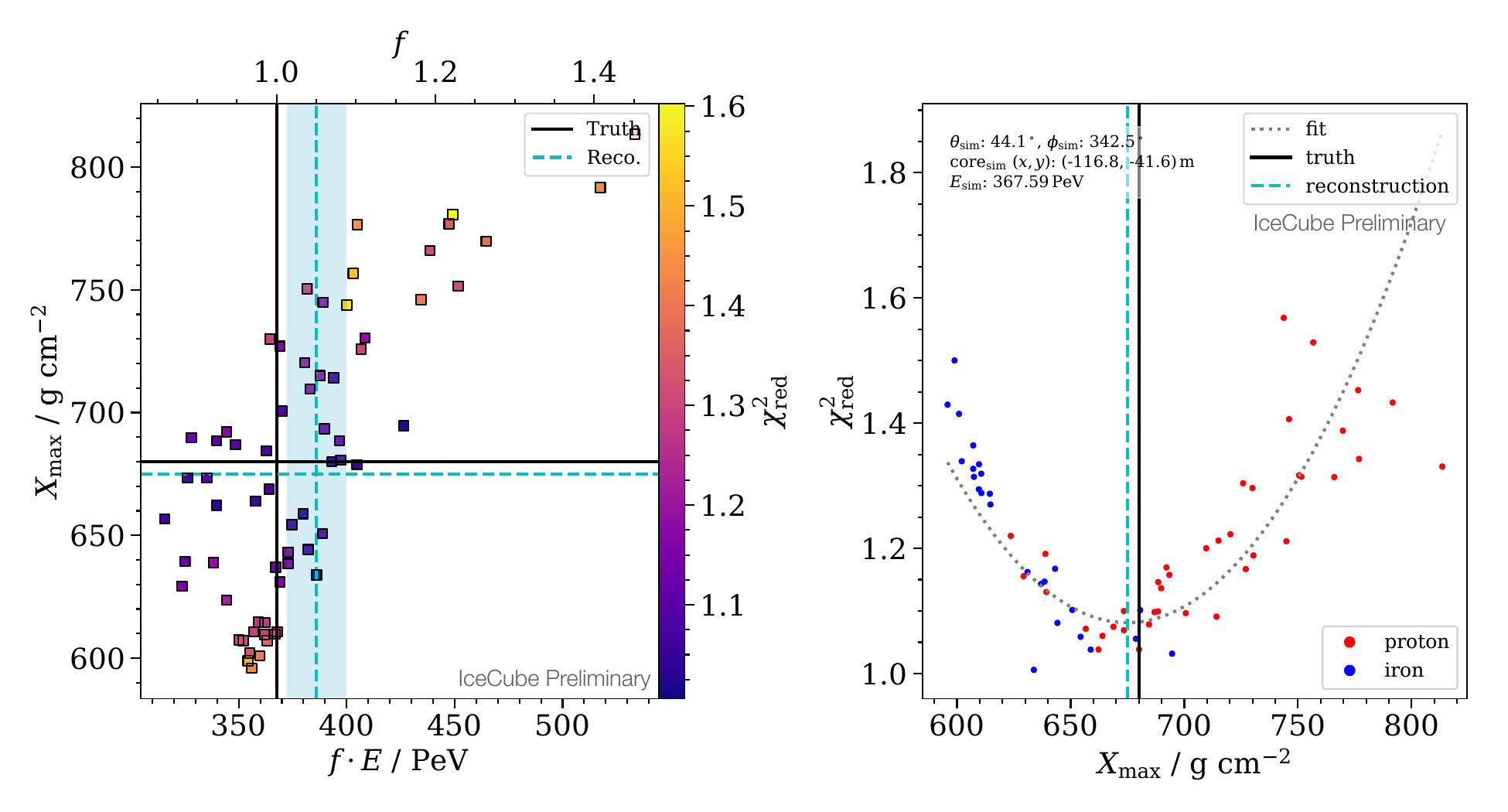}
    \caption{\xmax{} reconstruction of a mock data event with real noise injected, and filtered with the median filter. The characteristic parabola can be seen as well as the true and reconstructed values. The blue band represents the uncertainty on $f$ from the fitting procedure. }
    \label{fig:mockData_oneReco}
\end{figure}

An example of a reconstruction using mock data is presented in \reffig{fig:mockData_oneReco}. The plot illustrates the relationship between the \chisquared{} value obtained from the minimization process and \xmax{}. The \chisquared{} values form a parabolic shape, whose minimum is taken as the reconstructed value of \xmax{} ($X_\mathrm{max}^\mathrm{reco}$) for that event.

Each data point in the plot corresponds to the reconstruction of a mock air-shower event against one simulated event from its air-shower simulation set using the \refeq{eq:chi2} formula. To obtain the parabolic shape, a fitting procedure is applied to all data points except the highest 10 \chisquared-values.

\begin{figure}
    \centering
    \includegraphics[width=0.6\linewidth]{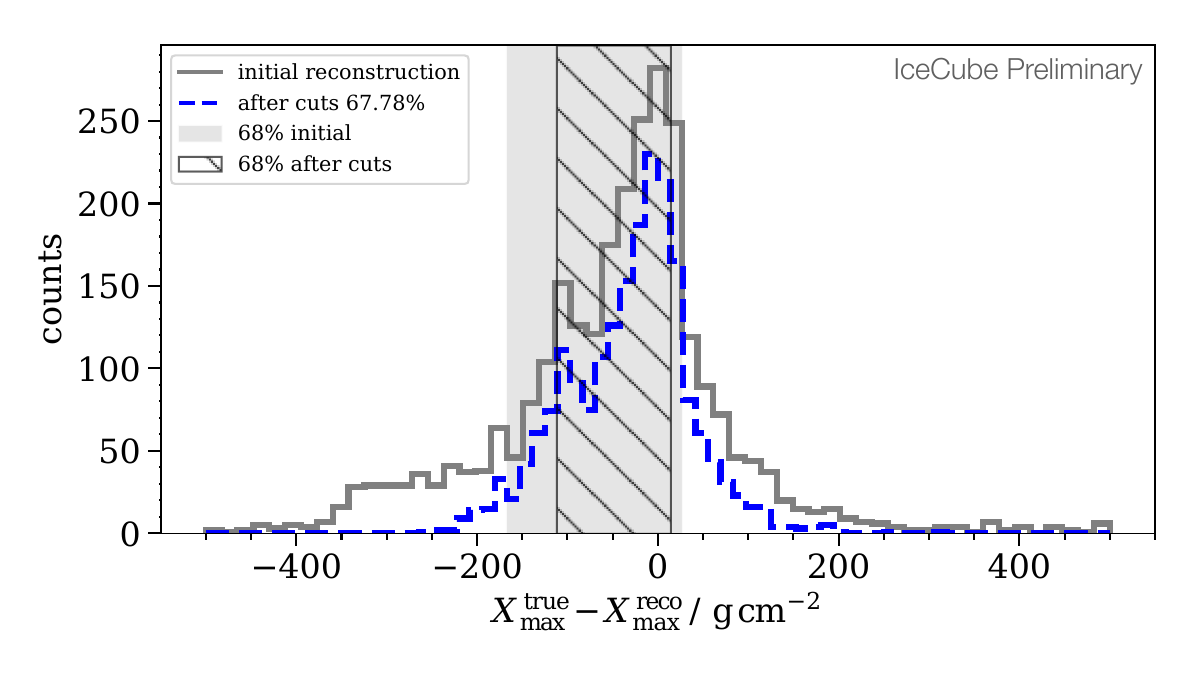}
    \caption{Histogram of all the reconstructions from Set 2 with the median filtering schemes before and after selection cuts. }
    \label{fig:histogram}
\end{figure}

Expanding on this principle, each simulation produced for one initial air shower is used as mock data. The reconstructed \xmax{} is then compared to the true \xmax{}($X_\mathrm{max}^\mathrm{true}$) of that simulation. The accuracy of the method is determined by calculating the difference between the reconstructed \xmax{} and the true \xmax{} for all simulations for one air shower, and then repeating this process for all air showers in a set.
The resulting histogram is depicted in \reffig{fig:histogram}.
For the dashed blue line, a few reconstruction quality cuts are applied; requiring a parabola with a positive quadratic coefficient and with its minimum within the range of the simulated \xmax{} values. The fraction of reconstructions meeting these criteria is specified in the legend.

\begin{figure}
    \centering
    \includegraphics[width=0.7\linewidth]{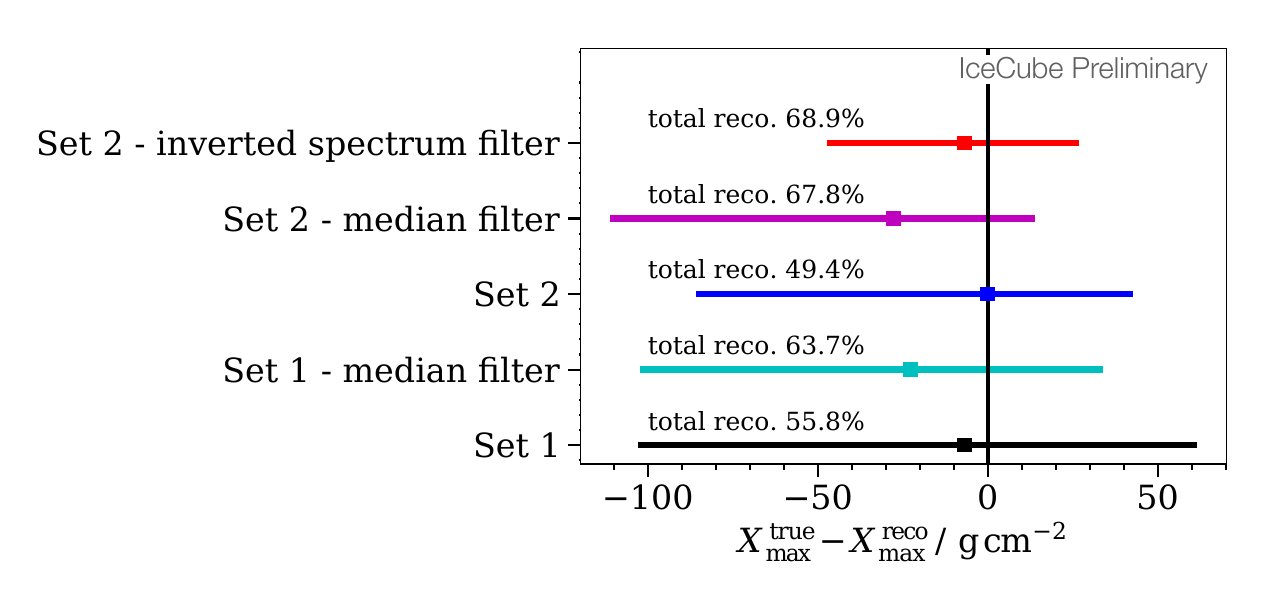}
    \caption{Summary of the resolution for the two different Sets of mock data using different filtering schemes. The error bars represent the 68\% containment of the reconstruction, and the total reconstruction percentage indicates the ratio of successful reconstructions out of all attempted reconstructions after applying the specified cuts.}
    \label{fig:comparing}
\end{figure}

This procedure is performed for both sets using different filtering techniques, as shown in \reffig{fig:comparing}. All waveforms are initially filtered with an 80-\SI{300}{\mega\Hz} Butterworth filter. Additional custom filters~\cite{ColemanICRC21} are then applied. The coined ``median filter'' involves smoothing the frequency spectrum of the specific waveform by calculating the sliding median in a 20-bin sliding window, aimed at reducing human-made noise peaks in the spectrum. The ``inverted spectrum filter'' pushes this idea further by creating an average spectrum from the TAXI 3.2 background data and dividing it by this spectrum after median frequency filtering. The resulting filter is applied to the waveforms twice, effectively acting as a notch filter designed based on the measured noise peaks. This filter is only applied to data of Set 2, as the mode of operation of Set 1 changed significantly throughout its run-time, making it difficult to establish a standard averaged spectrum for this data acquisition system.
 
\reffig{fig:comparing} compares the reconstruction efficiency for the two sets and demonstrates the impact of the different filtering schemes. 
Whereas the median filter results in a slightly better resolution compared to the standard 80-\SI{300}{\mega\hertz} filter, the inverted spectrum filter stands out as it significantly improves the reconstruction quality, with a resolution of $^{+27.0}_{-47.3}\,\si{\g\per\cm\squared}$ for the reconstructed \xmax{} with real noise injected.
Note that these resolution values assume an exact knowledge of the shower core and direction. The effect of the finite resolution in these quantities was not considered in the work presented here.

\section{Reconstruction applied on measured events}
\label{sec:data}
\begin{figure}
    \centering
    \includegraphics[width=0.95\linewidth]{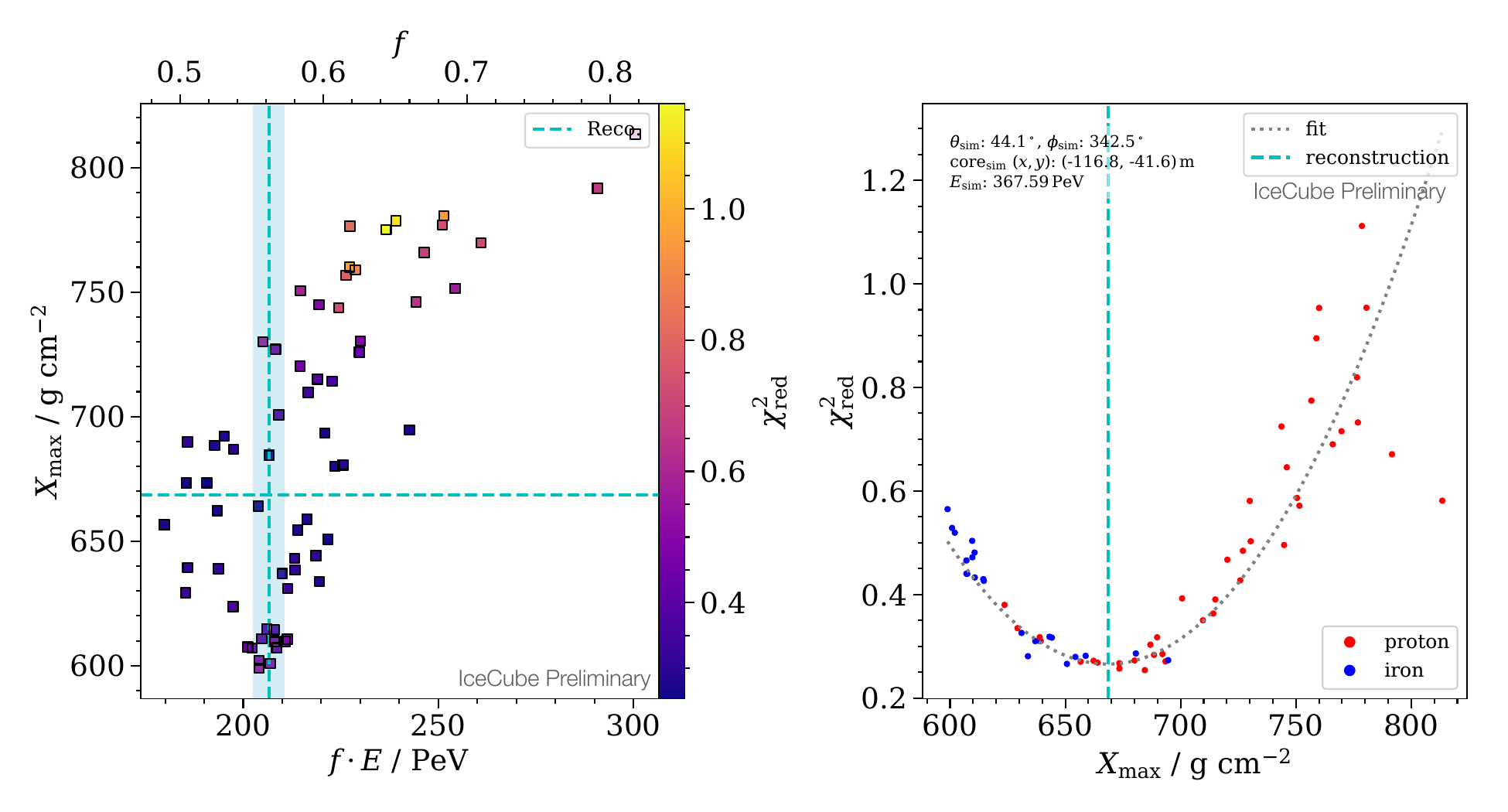}
    \caption{Reconstruction on measured data, the median filter was used.}
    \label{fig:data_reco}
\end{figure}

The technique is employed on the measured data, resulting in a reconstruction rate ranging from approximately 28\% to 49\%, depending on the Set and the filtering scheme. Successful reconstructions are defined by a reconstructed \xmax{} falling within the range of simulated \xmax{} values and exhibiting a positive parabolic fit. This rate is consistent with previous findings obtained from a smaller dataset~\cite{thesisRox, TurcotteARENA}.
An example of a reconstructed event using measured data is illustrated in \reffig{fig:data_reco}, which corresponds to the same air-shower event depicted in \reffig{fig:mockData_oneReco}. The parabolic shape is clearly discernible for this event.
The disparities observed between the reconstruction efficiency of real data and mock data are predominantly attributed to the uncertainties associated with calibration and core position determination.

\section{Discussion and conclusion}
We are currently implementing and refining the template-fitting \xmax{} reconstruction method to prepare for future deployments of the radio part of the surface array enhancement. However, there are certain challenges that need to be addressed. The method is primarily applied to real data out of curiosity and exploration, as we are aware that the calibration and core position determination are not yet accurate enough to enable the full reconstruction using only three antennas.

One of the main difficulties lies in the relatively small footprint covered by the three antennas compared to the footprint of the radiated energy from the air showers. This presents a challenge for the three-antenna reconstruction because it requires precise positioning within the footprint. Ideally, the antennas need to be located in a region where changes in the signal can be observed if \xmax{} is varied. For instance, if the antennas are located far from the Cherenkov ring, a change in energy or \xmax{} would have a similar effect on the signal observed by the antennas.

Despite these challenges, notable progress in employing improved filtering schemes is made. The ability to reconstruct \xmax{} for some events with only three antennas is remarkable. Furthermore, although the direct comparison between Sets has to be treated carefully due to the different event selections, the obtained result confirms the proper functioning of the TAXI 3.2. It also provides another confirmation for the correct identification of air-shower events with the prototype station and demonstrates the readiness for \xmax{} reconstruction with a larger array.

\bibliographystyle{ICRC}
\bibliography{references}

%

\clearpage

\section*{Full Author List: IceCube Collaboration}

\scriptsize
\noindent
R. Abbasi$^{17}$,
M. Ackermann$^{63}$,
J. Adams$^{18}$,
S. K. Agarwalla$^{40,\: 64}$,
J. A. Aguilar$^{12}$,
M. Ahlers$^{22}$,
J.M. Alameddine$^{23}$,
N. M. Amin$^{44}$,
K. Andeen$^{42}$,
G. Anton$^{26}$
% ...
\\
\\
% ...
$^{12}$ Universit{\'e} Libre de Bruxelles, Science Faculty CP230, B-1050 Brussels, Belgium \\
% ...
$^{17}$ Department of Physics, Loyola University Chicago, Chicago, IL 60660, USA \\
$^{18}$ Dept. of Physics and Astronomy, University of Canterbury, Private Bag 4800, Christchurch, New Zealand \\
% ...
$^{22}$ Niels Bohr Institute, University of Copenhagen, DK-2100 Copenhagen, Denmark \\
$^{23}$ Dept. of Physics, TU Dortmund University, D-44221 Dortmund, Germany \\
% ...
$^{26}$ Erlangen Centre for Astroparticle Physics, Friedrich-Alexander-Universit{\"a}t Erlangen-N{\"u}rnberg, D-91058 Erlangen, Germany \\
% ...
$^{40}$ Dept. of Physics and Wisconsin IceCube Particle Astrophysics Center, University of Wisconsin{\textendash}Madison, Madison, WI 53706, USA \\
% ...
$^{42}$ Department of Physics, Marquette University, Milwaukee, WI, 53201, USA \\
% ...
$^{44}$ Bartol Research Institute and Dept. of Physics and Astronomy, University of Delaware, Newark, DE 19716, USA \\
% ...
$^{63}$ Deutsches Elektronen-Synchrotron DESY, Platanenallee 6, 15738 Zeuthen, Germany  \\
$^{64}$ Institute of Physics, Sachivalaya Marg, Sainik School Post, Bhubaneswar 751005, India \\
% ...

\subsection*{Acknowledgements}

\noindent
The authors gratefully acknowledge the support from the following agencies and institutions:
USA {\textendash} U.S. National Science Foundation-Office of Polar Programs,
U.S. National Science Foundation-Physics Division,
U.S. National Science Foundation-EPSCoR,
Wisconsin Alumni Research Foundation,
Center for High Throughput Computing (CHTC) at the University of Wisconsin{\textendash}Madison,
Open Science Grid (OSG),
Advanced Cyberinfrastructure Coordination Ecosystem: Services {\&} Support (ACCESS),
Frontera computing project at the Texas Advanced Computing Center,
U.S. Department of Energy-National Energy Research Scientific Computing Center,
Particle astrophysics research computing center at the University of Maryland,
Institute for Cyber-Enabled Research at Michigan State University,
and Astroparticle physics computational facility at Marquette University;
Belgium {\textendash} Funds for Scientific Research (FRS-FNRS and FWO),
FWO Odysseus and Big Science programmes,
and Belgian Federal Science Policy Office (Belspo);
Germany {\textendash} Bundesministerium f{\"u}r Bildung und Forschung (BMBF),
Deutsche Forschungsgemeinschaft (DFG),
Helmholtz Alliance for Astroparticle Physics (HAP),
Initiative and Networking Fund of the Helmholtz Association,
Deutsches Elektronen Synchrotron (DESY),
and High Performance Computing cluster of the RWTH Aachen;
Sweden {\textendash} Swedish Research Council,
Swedish Polar Research Secretariat,
Swedish National Infrastructure for Computing (SNIC),
and Knut and Alice Wallenberg Foundation;
European Union {\textendash} EGI Advanced Computing for research;
Australia {\textendash} Australian Research Council;
Canada {\textendash} Natural Sciences and Engineering Research Council of Canada,
Calcul Qu{\'e}bec, Compute Ontario, Canada Foundation for Innovation, WestGrid, and Compute Canada;
Denmark {\textendash} Villum Fonden, Carlsberg Foundation, and European Commission;
New Zealand {\textendash} Marsden Fund;
Japan {\textendash} Japan Society for Promotion of Science (JSPS)
and Institute for Global Prominent Research (IGPR) of Chiba University;
Korea {\textendash} National Research Foundation of Korea (NRF);
Switzerland {\textendash} Swiss National Science Foundation (SNSF);
United Kingdom {\textendash} Department of Physics, University of Oxford.
This project has received funding from the European Research Council (ERC)  under the European Union’s Horizon 2020 research and innovation programme (grant agreement No 802729).

\end{document}